\documentclass[letterpaper,amsmath,amssymb,aps,prl,reprint,superscriptaddress,footinbib,floatfix]{revtex4-2}
\usepackage{geometry}
\usepackage{hyperref}
\usepackage{ulem}

\usepackage{graphicx}
\usepackage{bm}
\usepackage{amsmath,amssymb,framed,amsthm,xcolor} 
\usepackage{array,multirow,graphicx}
\usepackage{natbib}
\usepackage{float}

\usepackage{subcaption}
\usepackage{graphicx}
\usepackage{dcolumn}
\begin{document}
\title{Magnetic reconnection with a 0.1 rate:\\ Effective resistivity in general relativistic magnetohydrodynamics}

\author{B. Ripperda}
\altaffiliation[ripperda@cita.utoronto.ca]{}
\affiliation{Canadian Institute for Theoretical Astrophysics, 60 St. George St, Toronto, ON M5S 3H8, Canada}
\affiliation{Department of Physics, University of Toronto, 60 St. George St, Toronto, ON M5S 1A7, Canada}
\affiliation{David A. Dunlap Department of Astronomy \& Astrophysics, University of Toronto, 50 St. George St,Toronto, ON M5S 3H4, Canada}
\affiliation{Perimeter Institute for Theoretical Physics, 31 Caroline St. North, Waterloo, ON N2L 2Y5, Canada}

\author{M. P. Grehan}
\affiliation{Department of Physics, University of Toronto, 60 St. George St, Toronto, ON M5S 1A7, Canada}
\affiliation{Canadian Institute for Theoretical Astrophysics, 60 St. George St, Toronto, ON M5S 3H8, Canada}

\author{A. Moran}
\affiliation{Department of Astronomy, Columbia University,  New York, NY 10027, USA}

\author{S. Selvi}
\affiliation{Department of Astronomy, Columbia University,  New York, NY 10027, USA}

\author{L. Sironi}
\affiliation{Department of Astronomy, Columbia University,  New York, NY 10027, USA}
\affiliation{Center for Computational Astrophysics, Flatiron Institute, 162 Fifth Ave, New York, NY 10010, USA}

\author{A. Philippov}
\affiliation{Department of Physics, University of Maryland, 7901 Regents Drive, College Park, MD 20742, USA}

\author{A. Bransgrove}
\affiliation{Princeton Center for Theoretical Science and Department of Astrophysical Sciences, Princeton University, Princeton, NJ 08544, USA}

\author{O. Porth}
\affiliation{Anton Pannekoek Institute, Science Park 904, 1098 XH, Amsterdam, The Netherlands}


\begin{abstract}
Relativistic magnetic reconnection is thought to power various multi-wavelength emission signatures from neutron stars and black holes. Relativistic resistive magnetohydrodynamics (RRMHD) offers the simplest model of reconnection. However, a small uniform resistivity underestimates the reconnection rate compared to first-principles kinetic models. By employing an effective resistivity based on kinetic models --- which connects the reconnection electric field to the charge-starved current density --- we show that RRMHD can reproduce the increased reconnection rate of kinetic models, both in local current sheets and in global black hole magnetospheres.
\end{abstract}

\maketitle

\textit{Introduction.}---Relativistic magnetic reconnection is an efficient mechanism for converting magnetic into kinetic energy, powering multiwavelength radiation in high-energy astrophysical sources \cite{Sironi_2025,petropoulou_2016,Philippov_2019,most_20, lyubarsky_20,sironi_beloborodov_20, yuan_20,mahlmann_22,yuan_22, ripperda_22, crinquand_22, hakobyan_ripperda_23,Mbarek_2024,Fiorillo_2024,Most_2024,Selvi_2024}. 
The observed short and powerful flares from black holes and neutron stars require energy conversion to occur at a fast rate.

The simplest theoretical description of relativistic reconnection is provided by relativistic resistive magnetohydrodynamics (RRMHD).
In the Sweet-Parker (SP) model \cite{sweet_58,parker_57} and its relativistic extension \cite{lyutikov_uzdensky_03,lyubarsky_05}, a uniform resistivity $\eta$ results in a reconnection rate $\beta_{\rm{rec}}= v_{\rm in} / v_{\rm out} \sim S^{-1/2}$ which scales with the Lundquist number $S=4\pi v_{\rm A} \ell / \eta c^2$, 
the ratio between resistive and Alfv\'{e}nic time-scales for a layer of length $L=2\ell$, speed of light $c$, inflow speed into the layer $v_{\rm in}$, and outflow speed $v_{\rm out}$. When the magnetization $\sigma = B^2 / (4\pi n m c^2) \gg 1$, the outflow speed---of order the Alfv\'{e}n speed $v_{\rm A} = \sqrt{\sigma / (\sigma+1)}c$---approaches $c$, for a magnetic field strength $B$, total number density $n$ and particle mass $m$. Then $\beta_{\rm{rec}} \approx v_{\rm in}/c$, and hence $v_{\rm in}$ must be a significant fraction of $c$ to explain short flare time scales.

For realistically high astrophysical Lundquist numbers, SP predicts far too slow inflow velocity. However, in large $S \gtrsim 10^4$ plasma, current sheets can fragment into smaller reconnection layers, forming a chain of magnetic null points (X-points) and plasmoids. This plasmoid instability leads to an asymptotic rate $\beta_{\rm{rec}} \sim 10^{-2}$ \footnote{Derived specifically for non-relativistic incompressible resistive MHD, without Hall physics or other non-ideal effects, but confirmed to hold for small uniform resistivity in the relativistic regime \cite{ripperda_19b,Grehan_2025}}, faster than the SP result and independent of $\eta$ \cite{Bhattacharjee_2009,Uzdensky_2010}. 

In many relativistic astrophysical plasmas, the Coulomb mean free path is sufficiently large that RRMHD is formally inapplicable, and a kinetic treatment becomes necessary. In this collisionless regime, particle-in-cell (PIC) simulations show that reconnection proceeds at rates typically an order of magnitude higher than the asymptotic MHD value of $\sim 10^{-2}$ \cite{sironi_spitkovsky_14,guo_14,Werner_2015,sironi_16}. The faster rate is driven by anisotropic electron pressure localized in X-points, which has been suggested to produce an effect analogous to a spatially localized (i.e., non-uniform) resistivity \cite{bessho_05,selvi_23}.

For high $\sigma$, the large electric current in X-points combined with the low upstream density require the counterstreaming charged particles carrying the current to move at velocities approaching the speed of light. 
The plasma in the X-point cannot be replenished quickly enough and thus it becomes charge-starved. The resulting pressure and density deficits drive a fast inflow, i.e., a large $\beta_{\rm rec}$ \cite{goodbred_22}. Relating the charged starved current $|\bm{J}| \sim n e c$ for particles with charge $e$, to the non-ideal electric field in the X-point, \cite{Moran_2025} 
derived a guide-field independent and coordinate-agnostic effective resistivity. 

In this letter, we show that the effective resistivity derived by assuming that the current at X-points is charge starved is indeed essential to achieve the fast ``kinetic'' reconnection rate of $\beta_{\rm{rec}} \gtrsim 0.1$.  We demonstrate this via two-dimensional (2D) RRMHD simulations of a local Harris current sheet as well as global general relativistic resistive magnetohydrodynamics (GRRMHD) simulations of a split monopole black hole magnetosphere.





\begin{figure}
    \centering
    \includegraphics[width=\linewidth]{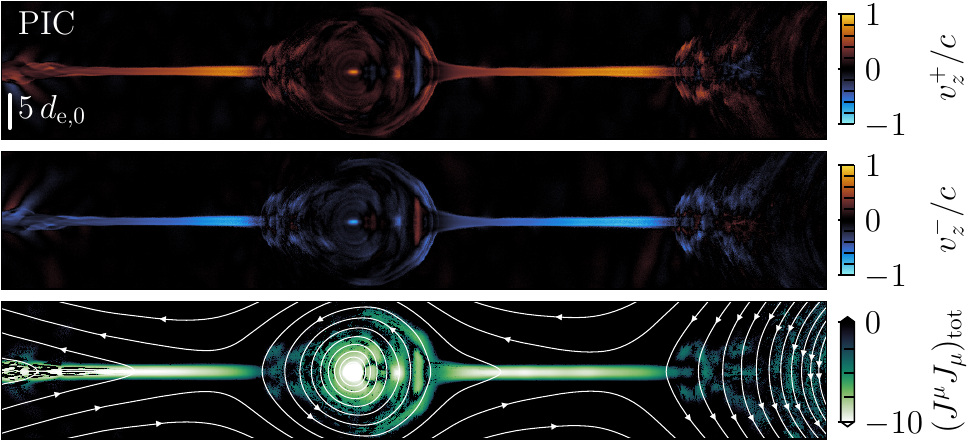}
    \caption{Charge starvation in a PIC simulation of a reconnecting Harris sheet. Positrons (first panel) and electrons (second panel) counter stream at near the speed of light. The Lorentz invariant $(J^\mu J_\mu)_{\rm tot}$ of the total current (third panel) remains spacelike ($<0$) in the current sheet showing that the current is supplied by counter streaming electrons and positrons. White lines and arrows show magnetic field lines linearly spaced in values of the out-of-plane vector potential. The third panel is normalized by $(n_0 e c)^2$ with upstream total number density $n_0$. The upstream skin depth $d_{e,0} = \sqrt{mc^2/4\pi n_0 e^2}$ is indicated by the vertical white line.}
    \label{fig:15}
\end{figure}

\textit{Resistivity Model.}---
In (G)RRMHD, the resistive source term in Amp\'{e}re's equation 
\begin{equation}
    \partial_t \bm{E} - c \nabla \times \bm{B} = 4\pi \bm{J}
\end{equation}
is given by Ohm's law $\bm{J} - q \bm{v} = \bm{E}^* / \eta$, where $ q=\nabla \cdot \bm{E}/4\pi$ is the charge density, $\bm{B}$ the magnetic field, $\bm{E}$ the electric field, $\bm{v}$ the velocity field with bulk Lorentz factor $\Gamma$, and 
\begin{equation}
    \bm{E}^* = \Gamma \left[\bm{E} + \bm{v} \times \bm{B}/c - (\bm{E}\cdot\bm{v}) \bm{v}/c^2 \right]
\end{equation}
\cite{Blackman_1993,komissarov_07}. Note that throughout this Letter we work exclusively in the lab frame.

\begin{figure}
    \centering
    \includegraphics[width=1.0\linewidth]{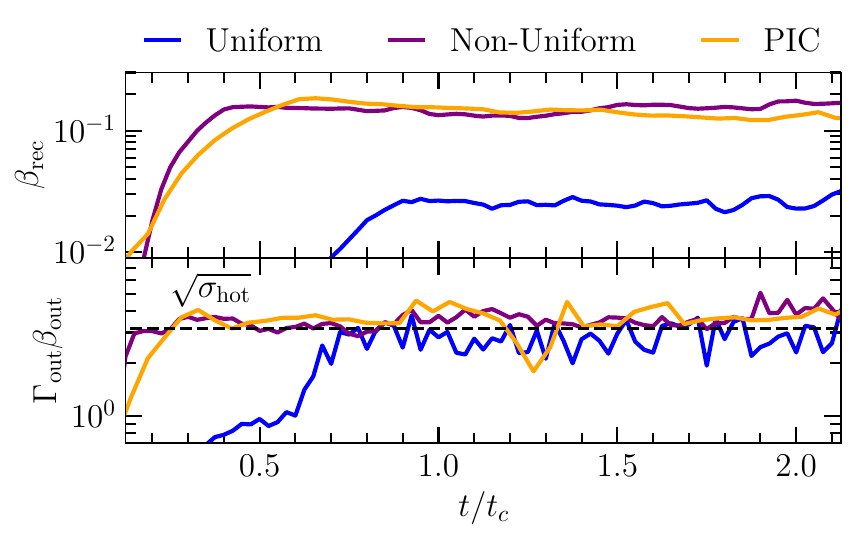}
    \caption{First panel: Time evolution of the reconnection rate $\beta_{\rm{rec}}$ for RRMHD with uniform $\eta$ (blue), non-uniform $\eta$ (purple), and PIC (orange), showing that the non-uniform $\eta$ case matches the PIC reconnection rate $\gtrsim 0.1$, while the uniform $\eta$ case shows the typical rate of RRMHD $\sim 0.03$. Second panel: Time evolution of the four velocity of the outflow out of the X-point (taken as the maximum in the domain), which becomes Alfv\'{e}nic as steady state is reached. The value expected for an Alfv\'{e}nic outflow ($\Gamma_{\rm A}\beta_{\rm A} = \sqrt{\sigma}$) is indicated by the black dashed line. 
    } 
    \label{fig:1}
\end{figure}

Charge starvation, and hence the effective $\eta$, is independent of the reference frame. This can be seen from a PIC simulation of a reconnection layer (described below), showing that the Lorentz invariant four-current magnitude for each particle species, $(J^{\mu} J_{\mu})_{\pm}= (n_{\pm} ec)^2 - (J^i J_i)_\pm \rightarrow0$ in the X-point, and hence each species drifts along (or opposite to) the direction of the electric current at nearly the speed of light
(see first and second panel in Fig. \ref{fig:15} for electron and positron velocities $v_z-\rightarrow -c$ and $v_z^+\rightarrow c$, respectively, where the current is in the out-of-plane $z$-direction). Equivalently, the Lorentz factor of the two counterstreaming species is much larger than unity at X-points.
The (Lorentz invariant) magnitude of the total plasma current is spacelike in the current layer indicating that indeed electrons and positrons have nearly equal densities and are couterstreaming (i.e., $(J^{\mu} J_{\mu})_{\rm tot}= (n_+ - n_-)^2 e^2c^2 - (J^i J_i)_{\rm tot} < 0$, third panel in Fig. \ref{fig:15}).  Due to the opposite velocities of the two species, the bulk velocity $\bm{v} \rightarrow 0$ and the bulk Lorentz factor $\Gamma\sim 1$ in the X-point. Note that the bulk Lorentz factor is large outside the X-point, reaching in the outflow $\Gamma \sim \sqrt{\sigma+1}$ due to $v_{\rm out} \sim v_{\rm A}$ (\cite{sironi_16,Sironi_2025}, see also Fig. \ref{fig:1}, second panel). 
The resistivity then follows from imposing charge starvation $|\bm{J}| \sim nec$ and $q\bm{v} \rightarrow 0$ in Ohm's law \cite{Moran_2025} and can be written in terms of single-fluid variables in the lab frame, namely (conserved) mass density $n m = \Gamma \rho$ with total density $n =n_+  + n_- $, $\bm{B}$, $\bm{v}$, and $\bm{E}$, which are readily available in our RRMHD scheme \cite{ripperda_19}:
\begin{equation}
\eta = \frac{|\bm{E}^*|}{n e c}.
\label{eq:1}
\end{equation}
We implicitly solve the source term in Amp\'{e}re's equation self-consistently for the electric field with an implicit-explicit Runge Kutta scheme \cite{ripperda_19}, where the resistivity (\ref{eq:1}) is calculated for each cell at each time. We set a small base resistivity such that the upstream Lundquist number $S_{\rm up} = 10^8$, so there is no division by zero in Ohm's law anywhere in the domain, while maintaining a near-ideal RMHD solution outside X-points.

The lab frame electric field $\bm{E}$ is dominated by its non-ideal component in the X-point (by definition, the fluid frame electric field \footnote{Note $\bm{E}^*$ differs minimally from the fluid frame electric field $\bm{E}' = \Gamma[\bm{E} + \bm{v}/c\times\bm{B} - \Gamma/(\Gamma+1)  (\bm{E} \cdot \bm{v})\bm{v}/c^2]$ \cite{Blackman_1993,komissarov_07}.}), such that the effective resistivity is localized in charge-starved regions with strong non-ideal electric fields, as suggested by kinetic simulations \cite{Moran_2025}.

In contrast to earlier works \cite{bessho_05,selvi_23,Bugli_2025}, this effective resistivity does not depend on derivatives, is coordinate agnostic, and is based on local quantities readily available in RRMHD.
Note that, unlike the resistivity model from  \cite{bessho_05,selvi_23,Bugli_2025}, our argument (and so, the form of $\eta$ in Equation \ref{eq:1}) also holds in the presence of a non-zero magnetic field component that does not reconnect (i.e., a guide field), because X-points are charge-starved and have $\bm{v} \rightarrow 0$ for all guide-field strengths \cite{Moran_2025}.
This makes our model generically applicable when it is not known a priori where the reconnection layer forms e.g., in relativistic magnetospheric, jet, or disk simulations \cite[e.g.,][]{tchekhovskoy_spitkovsky_13,ripperda_20,bransgrove_21,ripperda_22,Selvi_2024,Most_2024,Kim_2025} or relativistic turbulent plasmas \cite[e.g.,][]{Takamoto_16,bromberg_19,Chernoglazov_21}.

%

\begin{figure*}
    \centering
    \includegraphics[width=\linewidth]{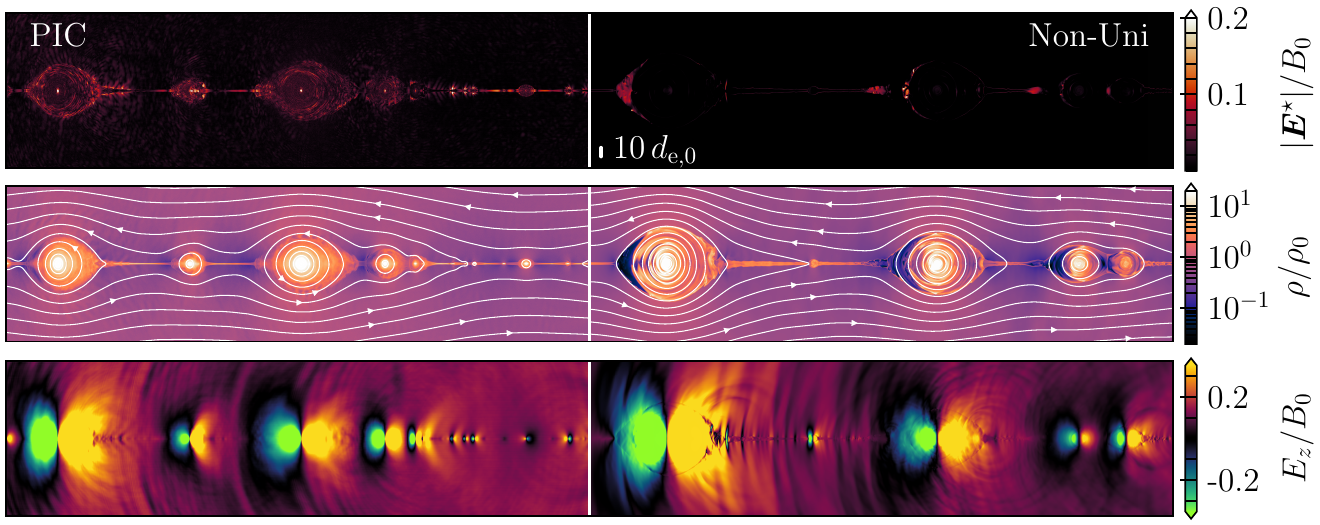}
    \caption{Comparison between PIC (left) and RRMHD with non-uniform $\eta$ (right) for $|\bm{E}^*|/B_0$ (top), $\rho/\rho_0$ (middle), and total out-of-plane electric field as a proxy for the reconnection rate $\beta_{\rm rec} \sim E_z / B_0$ (bottom). The X-points show a peak in $|\bm{E}^*|$ and drop in $\rho$, resulting in a fast $\beta_{\rm rec} \gtrsim 0.1$, both in RRMHD and PIC. White lines show contours of the out-of-plane vector potential with arrows indicating the direction of the in-plane magnetic field.
    }
    \label{fig:2}
\end{figure*}

\begin{figure}
    \centering
    \includegraphics[width=\linewidth]{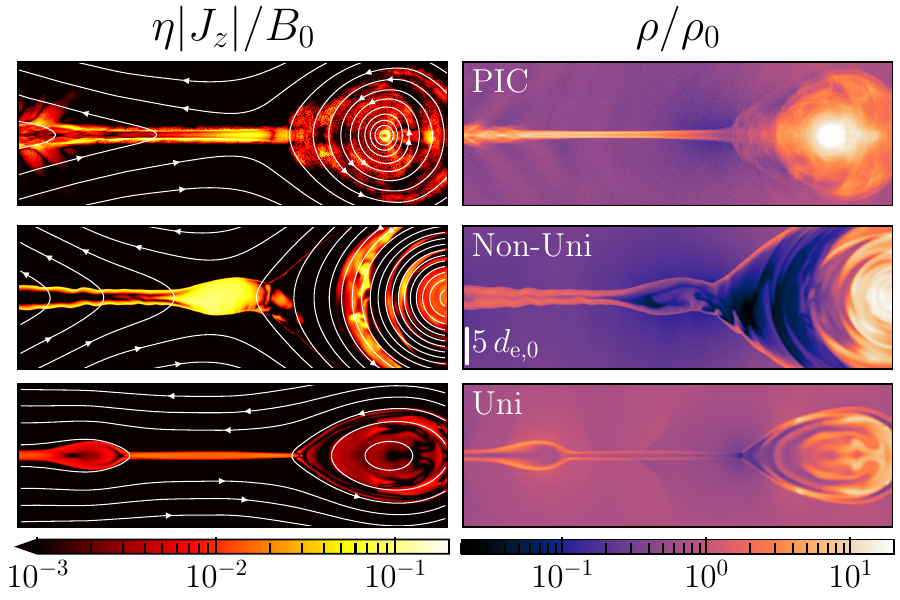}
    \caption{Zoom-in of non-ideal electric field component $\eta |J_z|/B_0$ (left) and density $\rho/\rho_0$ (right) in X-points for PIC (top), RRMHD with non-uniform $\eta$ (middle), and uniform $\eta$ (bottom). For uniform $\eta$, X-points are thinner and plasmoids are smaller.
    White lines show contours of the out-of-plane vector potential with arrows indicating the direction of the in-plane magnetic field.
    }
    \label{fig:3}
\end{figure}

\begin{figure*}
    \centering
    \includegraphics[width=\linewidth]{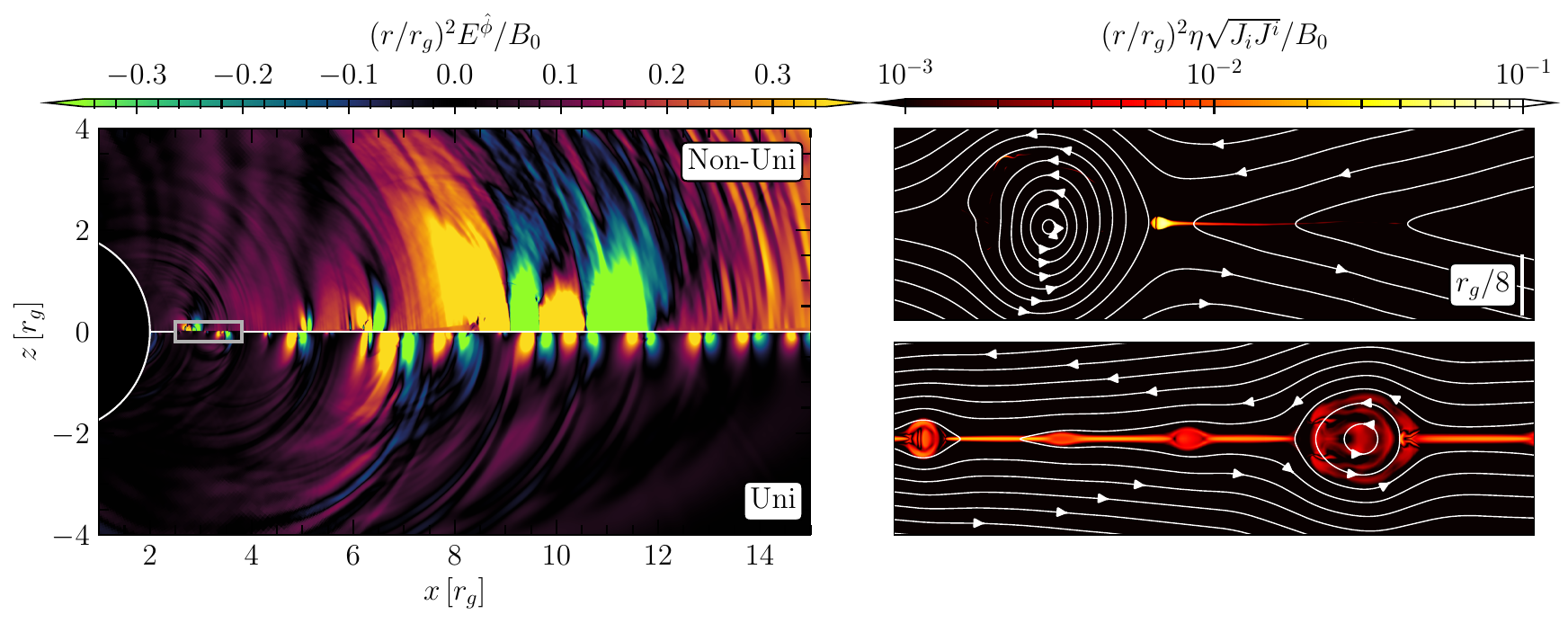}
    \caption{Left: The magnitude of the physical component of the reconnection electric field $(r/r_g)^2 {E}^{\hat{\phi}}/ {B}_0$ in the tetrad frame, indicating a reconnection rate $\beta_{\rm rec}\gtrsim 0.1$  for non-uniform $\eta$ (top half), and $\beta_{\rm rec} \sim 0.03$ for uniform $\eta$ (bottom half), where the white line divides the non-uniform and uniform case along the equator. 
    Right: The localization of the non-ideal electric field $(r/r_g^2) \eta \sqrt{J_{i} J^{i}}/{B}_0 \sim \beta_{\rm rec}$ is shown in a zoom in of the reconnection layer (top for non-uniform $\eta$, bottom for uniform $\eta$) as indicated by the silver rectangle on the left. White lines show contours of the out-of-plane vector potential, with arrows indicating the direction of the in-plane magnetic field. Both quantities are compensated by a factor $(r/r_g)^2$ to account for the decay of the monopolar upstream magnetic field component ${B}^{\hat{r}} \propto (r_g/r)^2$. 
    }
    \label{fig:4}
\end{figure*}

\textit{Numerical setups.}---To show that the resistivity model results in a $\sim 0.1$ reconnection rate, we simulate a 2D Harris current sheet \cite{harris_62} in Cartesian Minkowski spacetime $(t,x,y,z)$ in RRMHD with the Black Hole Accretion Code (BHAC, \citep{BHAC,Olivares2019,ripperda2019b}), as well as a 2D Harris current sheet in a pair plasma in PIC with TRISTAN-MP \cite{Tristan}. 

The current sheet is set by initializing a reconnecting magnetic field $B_x = B_0 \tanh ({y}/{w})$, and a guide field $B_z/B_0 = B_{\rm g}/B_0 \in [0,0.3,1]$ (see Supplemental Material for non-zero guide field cases). The initial velocity and electric fields are zero. We set a density profile $\rho = \rho_{0} + \rho_{\rm s} / \cosh^2(y/w)$ with over-density in the sheet $\rho_{\rm s} = 3 \rho_{0}$, upstream density $\rho_0$ set by the upstream $\sigma_0 = B_0^2 / (4\pi \rho_0 c^2) = 10$ and initial sheet thickness $w=2\times 10^{-2}L$ for box length $L$. Force balance determines the pressure profile $p = p_{0} + B_0^2 /(2\cosh^2(y/w ))$ with upstream pressure $p_0$ set by the dimensionless upstream temperature $p_{0}/\rho_{0}c^2=10^{-2}$.  We initialize reconnection by multiplying the spatially dependent part of the pressure by a small perturbation with amplitude $0.15$ localized at the center of the sheet \cite{Grehan_2025}. Both PIC and RRMHD simulations employ outflow boundaries such that waves and plasmoids can escape the box. In this way, we ensure that reconnection reaches a steady state.

The non-uniform resistivity effectively introduces an electron inertial length, or cold upstream skin depth  $d_{\rm e,0} = \sqrt{m c^2 / 4 \pi n_0 e^2}$. 
One can then use Amp\'{e}re's law and the charge-starvation condition, yielding $4\pi|\bm{J}|/c=|\nabla \times \bm{B}| \sim B_0/\Delta_0 \sim 4\pi n_0 e$ to estimate the current sheet thickness as
$\Delta_0 = B_0/ (4\pi n_0 e) = d_{\rm e,0} \sqrt{\sigma_0}$.
We parameterize equation \ref{eq:1} in terms of $\Delta_0$ such that 
\begin{equation}
    \eta =  4\pi t_c   \left(\frac{\Delta_0}{L}\right)  \left(\frac{|\bm{E}^*|}{B_0}\right) \left(\frac{n_0}{n}\right),
\end{equation}
and $t_c$ is the light crossing time per box length $L$. For RRMHD with effective $\eta$, we set a fiducial $\Delta_0/L = 10^{-3}$, and we confirm in the Supplemental Material that the results are independent of the effective skin depth for $\Delta_0 \ll L$, as in PIC \cite{Sironi_2025}. For PIC we set $\Delta_0/L = 3\times 10^{-3}$.
For non-uniform $\eta$, we find converged results (see Supplemental Material) with similar requirements as in PIC \cite{Sironi_2014}, i.e., we need $\gtrsim 5$ cells per $\Delta_{0}$ and $\Delta_{0} \lesssim L/100$.
For uniform $\eta$ we set $S= 2 \times 10^5$, such that it is well above the plasmoid instability limit \cite{ripperda_19b,Grehan_2025}. 

We also simulate a current sheet in an axisymmetric split monopole black hole magnetosphere in logarithmic Kerr-Schild coordinates ($t,r,\theta,\phi$) in GRRMHD \cite{BHAC}, with gravitational length scale $r_g = GM/c^2$, gravitational constant $G$, black hole mass $M$, zero black hole spin, and uniform magnetization $\sigma=25$ \cite{Selvi_25}. We set the resistivity in Equation \ref{eq:1} such that the expected current sheet thickness at the horizon is
$\Delta_{\rm 0,h} = d_{e,h} \sqrt{\sigma} \sim 10^{-3}\, r_g$, and it is resolved with 20 cells. We compare to a case with uniform resistivity corresponding to a Lundquist number $S = 10^5$ as defined using using fiducial current sheet length $L=10 r_g$.

\textit{Results: Local --- Harris current sheet.}---Fig. \ref{fig:1} shows that the steady state reconnection rate for the non-uniform $\eta$ model, $\beta_{\rm{rec}} \sim 0.14$ (purple line), is similar to the PIC result (orange line), and a factor $>4$ larger than the uniform $\eta$ result (blue line). The second panel shows that the bulk outflow speed is similar as well between PIC and the non-uniform $\eta$ case and matches the analytic Alfv\'{e}nic expectation $\Gamma_{\rm out} \beta_{\rm out} \sim \sqrt{\sigma}$ \cite{lyubarsky_05,Grehan_2025,Sironi_2025}, while it is slightly larger than for the uniform $\eta$ case. 

Fig. \ref{fig:2} shows that the non-uniform $\eta$ simulation  (right) reproduces the PIC result (left), with the electric field $|\bm{E}^*|/B_0$ peaking in the X-points (top panel). Note that $\bm{E}^*$ is more variable in PIC outside the X-points, but remains orders of magnitude smaller than in X-points. The density (middle panel) shows enhancement in plasmoids and at the edges of X-points, and deficit in the center of X-points (leading to charge starvation) compared to the upstream density ($\rho/\rho_0  \sim 0.2$ at X-points for the non-uniform $\eta$ case, see Supplemental Material), indicating that compressibility effects may play a role \cite{Parker1963}. The bottom panels show the reconnection rate $E_z/B_0 = \beta_{\rm{rec}} \gtrsim 0.1$. In the Supplemental Material we show that these conclusions hold for a range of $\Delta_0$, guide fields, and we further comment on the comparison to uniform $\eta$.

In Fig. \ref{fig:3} we show the non-ideal electric field component $\eta |J_z| / B_0$ (left) and density (right) near an X-point for PIC (top), non-uniform $\eta$ (middle) and uniform $\eta$ (bottom). 
The non-uniform $\eta$ and PIC panels have an equal number of skin depths across the plotted domain, whereas the uniform $\eta$ panel has no associated skin depth,
and we chose the same box length (in units of $L$) as for the non-uniform $\eta$ panel.
The non-uniform resistivity in the X-point is a function of $\Delta_0$ with $(\eta/4\pi)/(\Delta_0/c) \sim 0.4$: for smaller $\Delta_0$, the maximum current ${\rm{max}}(|J_z|) \sim \rm{max}$$(|\nabla \times \bm{B}|_z) \sim B_0/\Delta_0$ increases, and $\eta$ decreases to maintain a non-ideal electric field and reconnection rate of $E_z/B_0 \sim \eta |J_z| / B_0 \sim \beta_{\rm rec} \gtrsim 0.1$ (see Supplemental Material for a detailed analysis). 
The non-ideal electric field is an order of magnitude smaller in the uniform $\eta$ case resulting in a slower reconnection rate. The uniform $\eta$ case also results in a strong non-ideal field inside plasmoids, unlike in the PIC and non-uniform $\eta$ cases. 
The non-uniform $\eta$ case shows a similar X-point and plasmoid structure as the PIC result. The X-point in the middle row has a thickness $\sim \Delta_0 \sim 3 d_{e,0}$, and aspect ratio $a=L_{\rm X} / \Delta_0 \sim 4$, where $L_{\rm X}$ is the length of the X-point along $x$. The X-point is much thinner (with aspect ratio $\sim{\eta c/4\pi L} \sim 1/S$) in the uniform $\eta$ case (bottom panels), requiring higher grid resolution to be properly resolved (see Supplemental Material).




\textit{Results: Global --- Current sheet in a split-monopole black hole magnetosphere.}---To show the wide range of applicability of the effective resistivity in a global simulation, we simulate a current sheet in a split monopole black hole magnetosphere with 2D (axisymmetric) GRRMHD, as an exemplary model of reconnection powering a flare from near the event horizon of a supermassive black hole \cite{ripperda_20,ripperda_22} or a post-merger or post-collapse remnant black hole \cite{bransgrove_21,Selvi_2024,Most_2024,Kim_2025}. 

The rate at which flux is expelled from the event horizon is set by the reconnection rate \cite{bransgrove_21,Selvi_25}. Fig. \ref{fig:4} shows the reconnection electric field which sets the inflow speed into the equatorial current sheet $(r/r_g)^2 {E}^{\hat{\phi}}/{B}_0 \sim \beta_{\rm rec}$ in the tetrad frame [e.g., \cite{White_2016}], where $B_0$ is the upstream magnetic field at the horizon.
The top half of the panel shows the non-uniform $\eta$ case resulting in $\beta_{\rm rec }\gtrsim 0.1$, and the bottom half shows the uniform $\eta$ case resulting in a slower $\sim 0.02$ rate. The zoomed-in panels on the right show the magnitude of the localized non-ideal electric field in the non-uniform $\eta$ case (top) and the smaller, and less localized, non-ideal electric field in the uniform $\eta$ case (bottom).

\begin{figure}
    \centering
    \includegraphics[width=\linewidth]{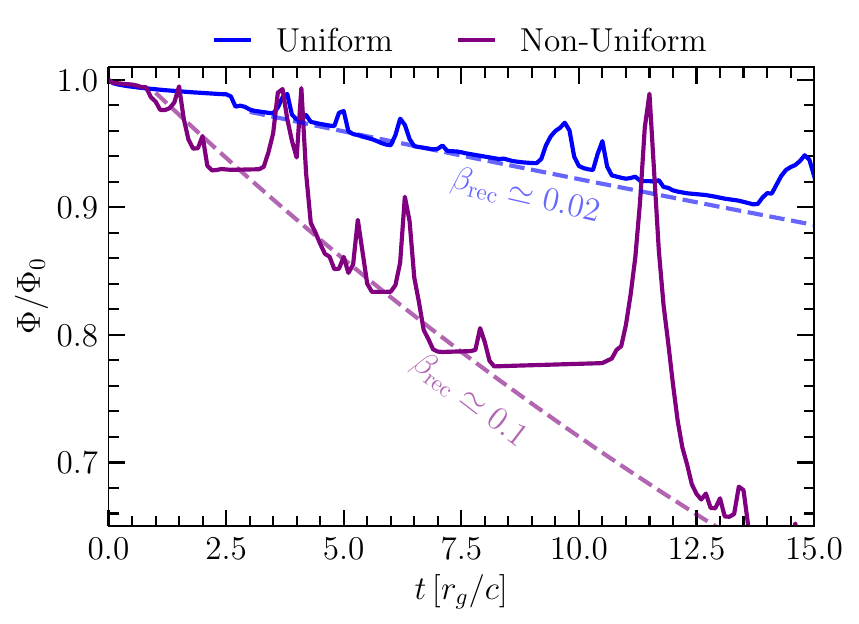}
    \caption{Time evolution of the normalized magnetic flux, $\Phi/\Phi_0$, at the event horizon, for uniform (blue), and non-uniform $\eta$ (purple). Dashed lines show the expected magnetic flux decay for fast and slow plasmoid-mediated reconnection (i.e., assuming $\beta_{\rm rec} \sim 0.1$ vs $0.02$ respectively \cite{bransgrove_21}) for a non-spinning black hole embedded in split-monopolar magnetic field \cite{bransgrove_21,Selvi_2024}.
    }
    \label{fig:5}
\end{figure}

The faster reconnection rate results in a faster magnetic flux $\Phi$ decay on the event horizon, as exemplified in Fig. \ref{fig:5}. The greater reconnection rate leads to bigger plasmoids, which give the larger spikes seen in Fig. \ref{fig:5} (one for every plasmoid falling through the horizon).
The lower bound of the flux decay matches with analytic expectations based on the reconnection rate (dashed lines, \cite{bransgrove_21,Selvi_2024,Selvi_25}), and in the non-uniform $\eta$ case it matches with the increased rate of general relativistic PIC simulations \cite{bransgrove_21,galishnikova_23}. 

\textit{Discussion.}---
The presented resistivity model can enable global 3D (G)RRMHD simulations of black hole or neutron star magnetospheres \cite{tchekhovskoy_spitkovsky_13}, jets \cite{bromberg_19}, coronae \cite{liska_22}, and accretion disks \cite{ripperda_22} as well as collapsing or merging compact objects \cite{Most_2024,Kim_2025}, capturing the physics of collisionless magnetic reconnection while achieving scale separations (e.g., with adaptive mesh refinement \cite{ripperda_19}) that are prohibitive in fully kinetic models \cite[e.g.,][]{Philippov_2019, bransgrove_21, galishnikova_23}. 

Due to the larger X-point size (as a result of the locally enhanced resistivity), the resolution requirement to resolve non-ideal regions in (G)RRMHD simulations with non-uniform resistivity is less stringent than for uniform resistivity \cite{ripperda_20} (see Supplemental Material).

The resistivity employed here is derived based on magnetically dominated electron-positron PIC simulations \cite{Moran_2025}, and the essential assumption of the charge-starved X-point may not hold for trans-relativistic or non-relativistic regimes where $\sigma \lesssim 1$ \cite{Robbins_25}. Additionally, if ions are present, Hall effects may be important. However, for electron-ion plasma with a large ion magnetization, the post-reconnection scale separation between species reduces due to efficient electron heating to ultra-relativistic temperatures (which then lowers the ratio of effective masses), suggesting that relativistic electron-ion reconnection behaves similarly as in pair plasma \cite{Sironi_2025}.

\section*{Acknowledgments}


In loving memory of Nuno Loureiro and all his inspiring contributions to magnetic reconnection and plasma physics.

The authors thank S. Komissarov, E. Most, A. Levis, D. Groselj, E. Quataert, A. Galishnikova, A. Savelli, J. Beattie, T. Ghosal, N. Murray, F. Bacchini, V. Berta, and C. Thompson for useful discussions. 

BR and MPG are supported by the Natural Sciences \& Engineering Research Council of Canada (NSERC) [funding reference number 568580], and the Canadian Space Agency (23JWGO2A01). 
MPG acknowledges the support of the Natural Sciences and Engineering Research Council of Canada (NSERC), [CGS D - 588952 - 2024]. Cette recherche a \'{e}t\'{e} financ\'{e}e par le Conseil de recherches en sciences naturelles et en g\'{e}nie du Canada (CRSNG), [CGS D - 588952 - 2024].
BR, LS and AP acknowledge support by a grant from the Simons Foundation (MP-SCMPS-00001470). This research was supported in part by grant NSF PHY-2309135 to the Kavli Institute for Theoretical Physics (KITP). 
LS acknowledges support from DoE Early Career Award DE-SC0023015, NASA ATP 80NSSC24K1238, NASA ATP 80NSSC24K1826, and NSF AST-2307202. AP additionally acknowledges support by NASA grant 80NSSC22K1054, Alfred P.~Sloan Research Fellowship and a Packard Foundation Fellowship in Science and Engineering. 
This research was facilitated by the Multimessenger Plasma Physics Center (MPPC), SS, LS and AP, acknowledge support from the NSF grants PHY-2206607 and PHY-2206609.
AB is supported by a PCTS fellowship and a Lyman Spitzer Jr. fellowship.
The computational resources and services used in this work were partially provided by facilities supported by the VSC (Flemish Supercomputer Center), funded by the Research Foundation Flanders (FWO) and the Flemish Government – department EWI, by Compute Ontario and the Digital Research Alliance of Canada (alliancecan.ca) compute allocation rrg-ripperda, and the CCA at the Flatiron Institute supported by the Simons Foundation.

\clearpage
\onecolumngrid

\section*{Supplemental Material}

In this Supplemental Material, we provide additional details and convergence tests of our simulations. We analyze the properties of a typical X-point to show that it locally resembles a compressible Sweet-Parker current sheet. We also present cases with a non-zero guide magnetic field.

\section{Convergence}

\begin{figure}
    \centering
        \includegraphics[width=0.7\linewidth]{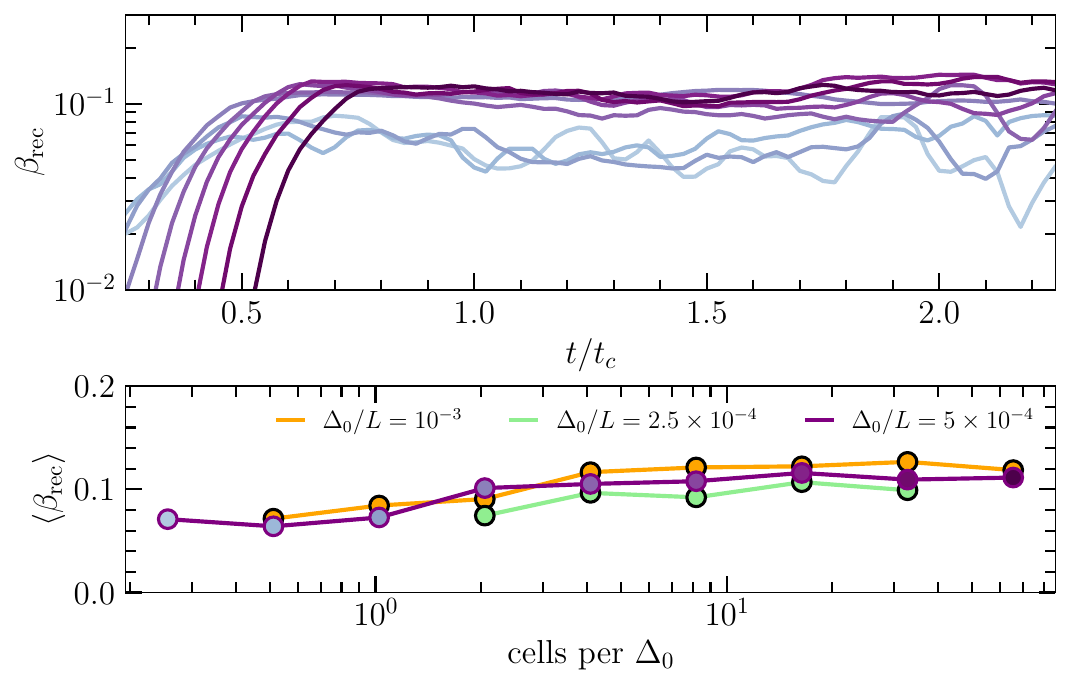}
    \caption{Top panel: Time evolution of the reconnection rate $\beta_{\rm rec}$ for a range of resolutions at fixed $\Delta_0/L=5 \times 10^{-4}$, corresponding to the different tones of purple in the second panel, indicating cells per effective skin depth $\Delta_0$. For resolution  $\gtrsim 4$ cells per $\Delta_0$ the reconnection rate converges (light to dark purple lines). For lower resolutions (light blue lines) the reconnection rate is smaller. Bottom panel: time-averaged reconnection rate, $\langle \beta_{\rm rec}\rangle$, in the steady-state stage, showing a converged $\beta_{\rm rec} \gtrsim 0.1$ for 5 cells per $\Delta_0$ for three different $\Delta_0/L \ll 0.01$.}
    \label{fig:amr_con}
\end{figure}

In Fig. \ref{fig:amr_con} we study convergence based on the reconnection rate by varying the numerical resolution (cells per sheet thickness $\Delta_0$). The effective hot skin depth $\Delta_0 = d_{\rm e,0} \sqrt{\sigma_0}$ sets the sheet thickness and is a parameter in our RRMHD simulations set by the  cold upstream magnetization $\sigma_0 = B^2 / (4\pi n_0 m c^2)$ and upstream density $\Gamma_0 \rho_0 =\rho_0= n_0 m$ for $\Gamma_0=1$ or effective cold upstream skin depth  $d_{\rm e,0} = \sqrt{m c^2 / 4 \pi n_0 e^2}$.
The top panel shows the reconnection rate $\beta_{\rm rec}$ for a fixed $\Delta_0/L=0.5 \times 10^{-3}$ where the color of the lines corresponds to the resolution (see second panel). The reconnection rate shows signs of convergence starting at approximately 2 cells per $\Delta_0$, and is converged at 4 cells per $\Delta_0$.

\begin{figure}
    \centering
        \includegraphics[width=\linewidth]{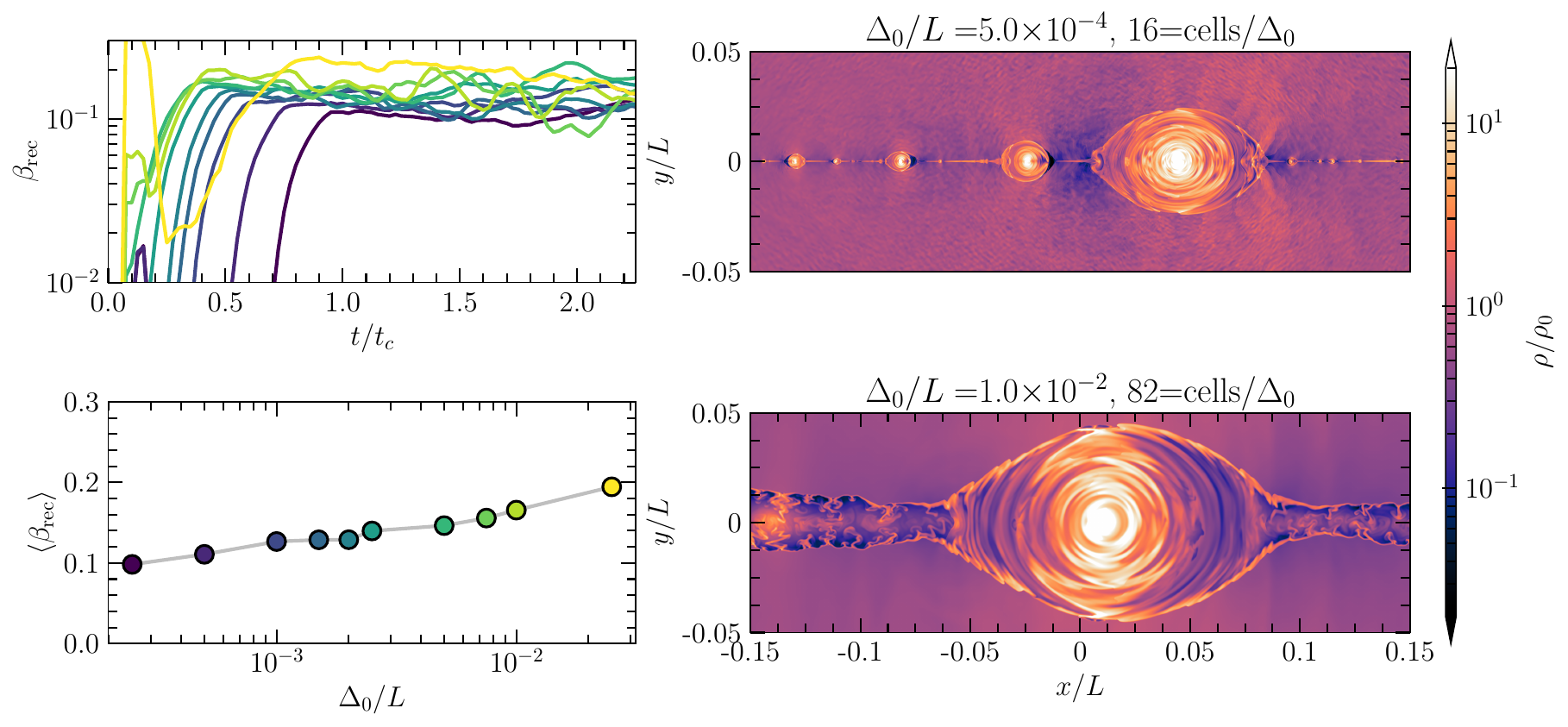}
    \caption{Reconnection rate as a function of effective skin depths per box length $\Delta_0/L$. Top left panel: The reconnection rate asymptotes to $\beta_{\rm rec} \gtrsim 0.1$ for all $\Delta_0/L$ (colors corresponding to $\Delta_0/L$ as in the bottom left panel). Current sheets in simulations with smaller $\Delta_0/L$ undergo tearing instability slightly later. Bottom left panel: Time-averaged reconnection rate, $\langle \beta_{\rm rec}\rangle$, in the steady state for a range of $\Delta_0/L$. The rate depends mildly on the $\Delta_0/L$, where for smaller $\Delta_0/L$ the larger number of plasmoids slightly inhibits the inflow. For effective skin depth $\Delta_0 \geq 10^{-2}L$, the reconnection rate is larger due to the current sheet thickness $\sim \Delta_0$ becoming a significant fraction of the box size (see bottom right panel). Top right panel: Steady-state reconnection with a large chain of plasmoids and X-points, where X-points as well as the smallest plasmoids have a width $\sim \Delta_0$. Bottom right panel: For $\Delta_0 \geq 10^{-2}L$, the current sheet thickness $\sim \Delta_0$, and hence the X-points become large with length $\sim 0.1 L$, resulting in an artificially thick current sheet with significant substructure.}
    \label{fig:delta_con}
\end{figure}

The bottom panel shows that the reconnection rate mildly depends on the ratio $\Delta_0/L$ (see also Fig. \ref{fig:delta_con}). Simulations with two and four times smaller $\Delta_0/L$ ratios (purple and green lines, respectively) compared to our fiducial $\Delta_0/L = 10^{-3}$, result in a well converged $\beta_{\rm rec} \gtrsim 0.1$ for  $\sim 10$ cells per $\Delta_0$, regardless of $\Delta_0/L$, in agreement with convergence studies for PIC simulations \cite{sironi_spitkovsky_14,Sironi_2025}.

In Fig. \ref{fig:delta_con} we show the convergence of the reconnection rate with $\Delta_0/L$ (top-left panel for $\beta_{\rm rec}$ as a function of time and bottom-left for time-averaged reconnection rates). We find a mild dependence on (a well-resolved) $\Delta_0/L$ resulting in $\beta_{\rm rec} \sim 0.1$ for the smallest $\Delta_0/L$, similar as in PIC simulations \cite{sironi_spitkovsky_14,Sironi_2025}. The marginal trend of lower $\beta_{\rm rec}$ for smaller $\Delta_0/L$ may be attributed to a larger number of plasmoids/magnetic islands forming at smaller $\Delta_0/L$, that can mildly inhibit the inflow, and hence the reconnection rate \cite{sironi_spitkovsky_14}.

The width of the current sheet is set by the effective hot skin depth $\Delta_0$ (right panels). For $\Delta_0/L < 0.01$ (upper right panel) we find a thin current sheet that is self-similar with $\Delta_0/L$ (i.e., smaller $\Delta_0/L$ results in a thinner current sheet, and the smallest plasmoids are born at smaller scales). If the effective skin depth is too large with respect to the box size, i.e., $\Delta_0 \gtrsim 10^{-2} L$ (lower right panel), then the current sheet becomes artificially thick, showing significant substructure (see also \cite{Bugli_2025}).

\begin{figure}
    \centering
    \includegraphics[width=0.65\linewidth]{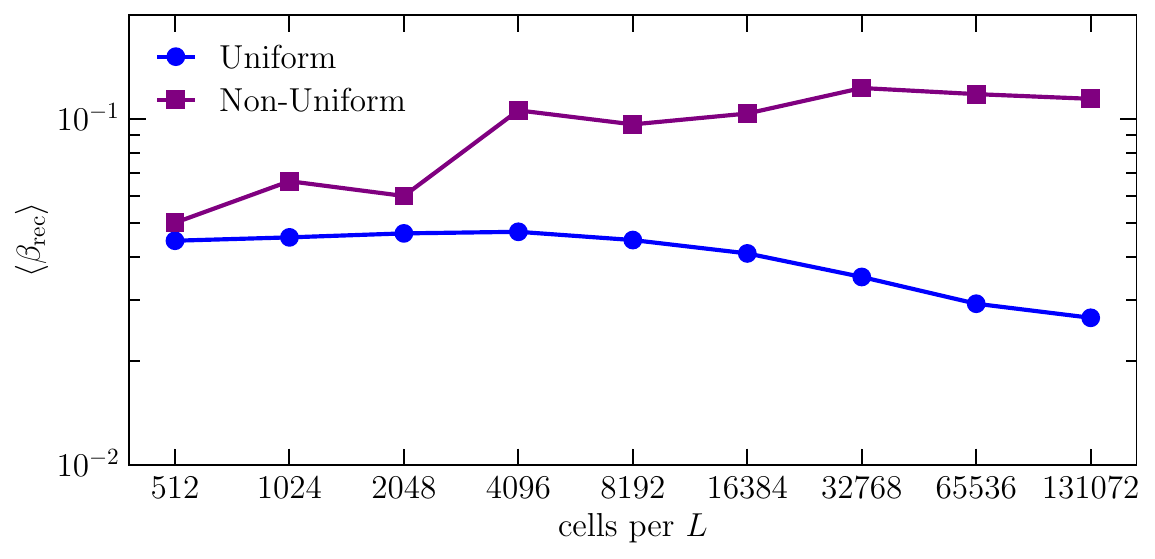}
    \caption{Time-averaged reconnection rate in the steady-state stage, versus resolution for non-uniform $\eta$ (purple line, for a fixed $\Delta_0/L=5\times 10^{-4}$, as in Fig. \ref{fig:amr_con}) and uniform $\eta$ corresponding to $S=2\times10^5$ (blue line). For non-uniform $\eta$ convergence with $\langle \beta_{\rm rec} \rangle \gtrsim 0.1$ is achieved for $\geq 4096$ cells per box (or 4 cells per $\Delta_0$), whereas the uniform $\eta$ case reaches the asymptotic rate in the plasmoid-unstable regime only with $\geq 65536$ cells per box (see also \cite{ripperda_19b,Grehan_2025}), over a factor of ten more.}
    \label{fig:uni_con}
\end{figure}

\begin{figure}
    \centering
    \includegraphics[width=0.85\linewidth]{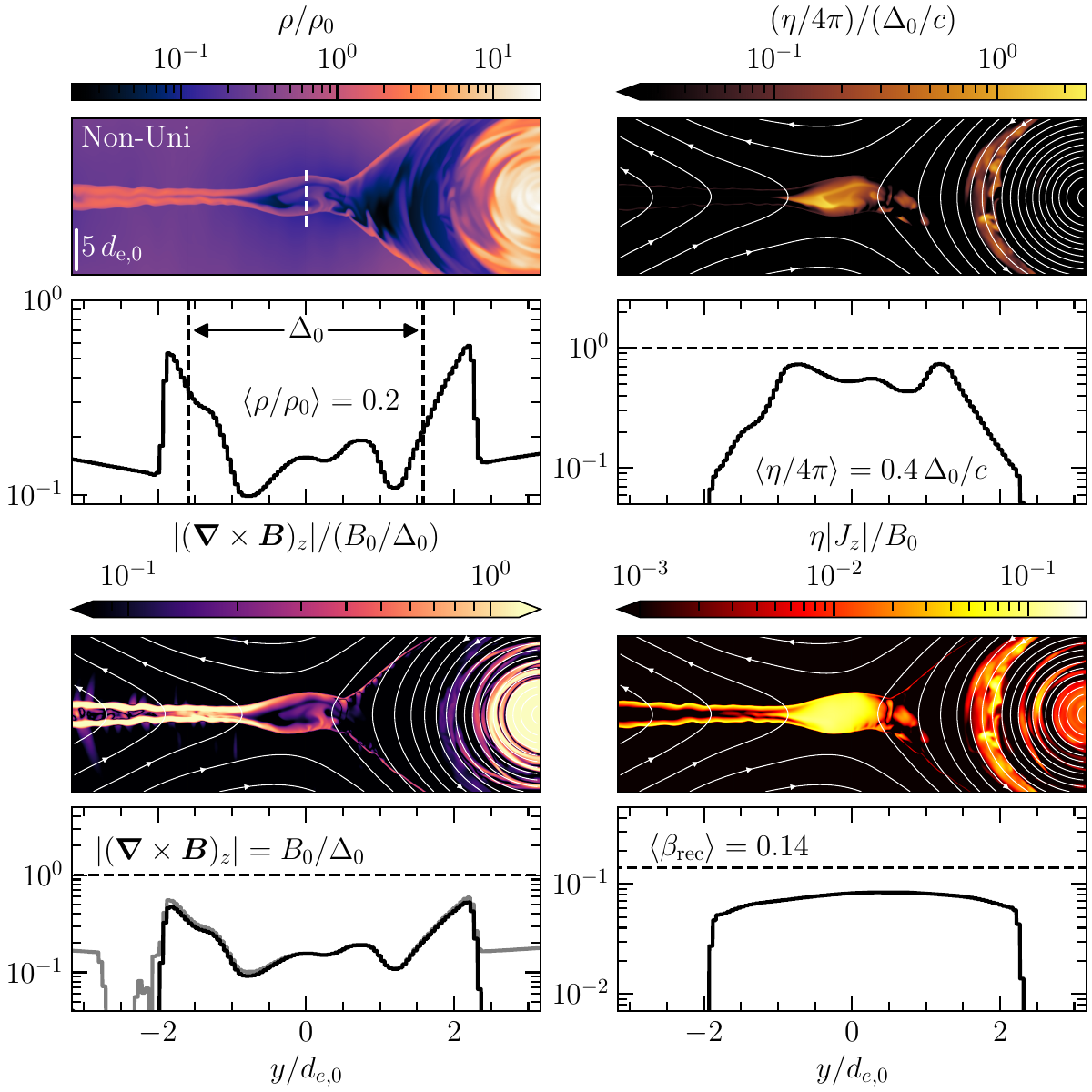}
    \caption{Analysis of an X-point for a case with non-uniform $\eta$ with fixed $\Delta_0/L=10^{-3}$. Panels in the second and fourth rows are 1D cuts taken along the white dashed line in the top left panel and correspond to the quantity in the panel above. 
    First column first and second row: The density in the X-point is $\sim 5$ times lower than the (far) upstream density $\rho_0$ (top panel) indicating a compressibility factor of $\sqrt{\rho_{0}/\rho_{\rm X}} \sim 2$ (bottom panel).
    Second column first and second row: Non-uniform resistivity in the X-point (top panel) showing that $(\eta/4\pi) / (\Delta_0/c) \sim 0.4$ (dashed line showing $(\eta/4\pi) = (\Delta_0/c)$, bottom panel).
    First column third and fourth row: The plasma conduction current $|\nabla \times \bm{B}|_z / (B_0/\Delta_0)$ peaks at the edges of the X-point (top panel), where it is approximately equal to $\sim B_0 / \Delta_0$ (dashed line, bottom panel).
    Second column third and fourth row: The non-ideal field component $\eta |J_z|/B_0$ is almost uniform in the X-point (top panel) and is $\sim \beta_{\rm rec}$ (dashed line, bottom panel).
    The average values for $\eta$ and $\rho$ are taken as spatial average in the 1D slices for all cells where $(\eta c/4\pi \Delta_0) > 10^{-1}$.
    White lines indicate linearly spaced contours of the out-of-plane vector potential and arrows show the direction of the in-plane magnetic field.
    }
    \label{fig:1d_slice}
\end{figure}

In Fig. \ref{fig:uni_con} we compare the time-averaged reconnection rate in steady state rate for a non-uniform resistivity with fixed $\Delta_0/L = 5 \times 10^{-4}$ and a uniform resistivity with fixed $S=2\times 10^5$, versus resolution. To compare with uniform $\eta$, we give resolution in terms of the total number of cells per box length $L$ because the uniform $\eta$ case does not have an associated effective skin depth. 
For non-uniform $\eta$ (purple line) we find convergence for resolution $N_x =4096$ along the current sheet, for a box of length $L$, which corresponds to $4$ cells per sheet thickness $\Delta_0/L=5\times 10^{-4}$ (similar as in Fig. \ref{fig:amr_con})
For uniform $\eta$ (blue line), which assumes a fixed $S=2\times10^5$, we find convergence for resolution $N_x = 65536 $ (see also \cite{ripperda_19b,Grehan_2025}), approximately 10 times higher (in terms of cells per $L$) than for non-uniform $\eta$ for the same box size $L$. The higher resolution requirement for uniform $\eta$ (specifically in the plasmoid-unstable asymptotic regime) is due to the larger aspect ratio of the X-points as a result of the the smaller reconnection rate, and hence, the fact that the region with non-ideal electric fields has smaller scale that needs to be resolved (see also Fig. \ref{fig:1d_slice}).

In both uniform and non-uniform cases, for smaller resolutions that do not resolve the sheet thickness set by the resistive length scale $\sim \eta c$ or effective skin depth scale $\Delta_0$ with $\gtrsim 4$ cells, the numerical resistivity $\eta_{\rm num} > \eta$ dominates \cite{Grehan_2025}. This results in a slightly higher reconnection rate $\beta_{\rm rec} \sim 0.04-0.05$ compared to the resolved rate in the asymptotic $S\gg 10^4$ regime, $\beta_{\rm rec} \sim 0.02-0.03$ for uniform $\eta$, due to the dominating, effectively smaller, numerical Lundquist number $S_{\rm num} < S$. For non-uniform $\eta$ the opposite effect is noticeable where the unresolved case results in a smaller $\beta_{\rm rec} \sim 0.04-0.05$ than the resolved $\beta_{\rm rec} \gtrsim 0.1$, due to dominating numerical resistivity $\eta_{\rm num} > \eta $. When numerical resistivity dominates (either due to unresolved resistivity, or in ideal MHD) there is no difference between an explicit uniform or non-uniform $\eta$, and the resulting reconnection rate is dominated by the uniform nature of the numerical resistivity (i.e., it is the same everywhere in the domain due to a uniform resolution in the current sheet) \cite{Grehan_2025}. For non-uniform resistivity simulations it is therefore particularly important to resolve the current sheet with $\geq 4$ cells per $\Delta_0$, which is however a less demanding criterion than for uniform resistivity due to the larger aspect ratio of X-points.

Note that $\eta |J_z| / B_0 \sim \beta_{\rm rec}$ is a good indicator of numerical convergence, since the current has reached its maximum value and is not limited by grid resolution (see also Fig. \ref{fig:1d_slice}). A $\beta_{\rm rec} > \eta |J_z| / B_0$ indicates that the reconnection is governed by numerical resistivity and the current sheet cannot reach its minimum thickness. In the current sheet in the split monopole black hole magnetosphere discussed in the main text, we once again find $\eta \sqrt{J_i J^i} / B_0 \sim \beta_{\rm rec}$, showing that the non-ideal field is properly resolved.

We note that a converged reconnection rate in time, i.e., a steady-state reconnection resulting in a time-averaged $\langle \beta_{\rm rec} \rangle \sim 0.1$ requires open boundaries along the outflow direction, such that plasmoids can leave the domain and do not artificially choke reconnection after a few box crossing times.


\section{Aspect ratio of X-points --- comparison to Sweet-Parker theory}

In this section we measure the aspect ratio, density, resistivity, current, and non-ideal electric field in a single X-point (for fixed fiducial $\Delta_0/L = 10^{-3}$ as in the main Letter). These estimates can then be used to test how well (incompressible) SP theory applies to an X-point, and how important the effect of compressibility is.

Fig. \ref{fig:1d_slice} shows zoom-ins and 1D cuts (along the white dashed line in the top-left panel) through the same X-point as in Fig. \ref{fig:3} in the main Letter for our fiducial $\Delta_0/L=10^{-3}$. White lines indicate contours of the out-of-plane vector potential, and white arrows the direction of the in-plane magnetic field, showing the magnetic geometry of the X-point.

The top two panels of the left column show the density $\rho/\rho_0$ in the X-point, indicating that the inner region is at lower density than the (far) upstream, and density peaks at the edges of the current sheet (top panel). Compressibility introduces a density difference between the upstream and the X-point $\sim {\langle \rho_{\rm X}/\rho_0 \rangle } \sim 0.2$ \cite{Parker1963,goodbred_22}. The thickness of the X-point $\sim \Delta_0 \sim 3 d_{e,0}$ is indicated by the vertical dashed lines (bottom panel), resulting in aspect ratio $a=L_{\rm X} / \Delta_0 \sim 4$. 
The average value for $\rho$ in the X-point is taken as spatial average in the 1D slices for all cells where $(\eta c/4\pi \Delta_0) > 10^{-1}$. We also measure $L_{\rm X}$ directly from the 1D slice as the length of the X-point (along $x$ where $(\eta c/4\pi \Delta_0) > 10^{-1}$). Note that both the averaging and the X-point length are insensitive to changes in the $(\eta c/4\pi \Delta_0)$ cutoff employed as the resistivity drops off rapidly outside the X-point.

The top two panels of the second column show that the resistivity peaks inside the X-point and is slightly smaller at its edges (top panel). The resistivity follows a scaling $\beta_{\rm rec} \propto \eta c / 4\pi \Delta_0$ (bottom panel, dashed line), where the proportionality constant depends on the compressibility factor of the X-point. We measure $(\eta/4\pi)/(\Delta_0/c) \sim 0.4$ (taken as an average in the X-point region), so the ratio of the local Lundquist number and aspect ratio is,  $S_{\rm X}/a = 4\pi v_{\rm A} \Delta_0 / \eta c^2 \sim 2.4$.

The bottom two panels of the first column show that the out-of-plane non-ideal current is equal to the curl of the magnetic field $|J_z| \sim |\nabla \times \bm{B}|_z$, and reaches a maximum value close to  $\sim B_0/\Delta_0$ (bottom panel, dashed line). The out-of-plane resistive current magnitude
\begin{equation}
|J_z| = \left|q v_z + \frac{\Gamma}{\eta} \left [E_z + (\bm{v} \times \bm{B})_z - (\bm{E} \cdot \bm{v}) v_z \right] \right|,    
\end{equation}
is indicated by the gray line and closely overlaps with $|\nabla \times \bm{B}|_z$ (the black line), showing that the displacement current is negligible compared to the plasma conduction current in the X-point. The top panel shows that the current peaks on the edges of the X-point.

The bottom two panels of the second column show that the non-ideal electric field $\eta |J_z| / B_0 \sim \beta_{\rm rec}$ (bottom panel), and it is uniform in the X-point (top panel). The uniformity of $\eta |J_z| / B_0$ suggests that for smaller $\Delta_0$, $|J_z| \sim |\nabla \times \bm{B}|_z \sim B_0/\Delta_0$ goes up, and $\eta$ has to go down to maintain $\eta |J_z|/B_0 \sim \beta_{\rm rec}$ (with the time-averaged $\langle \beta_{\rm rec} \rangle =0.14$ indicated by the horizontal dashed line).

Based on the measured aspect ratio and uniform $\eta |J_z| / B_0$, the X-point can be considered as an incompressible ``local'' SP current sheet with local Lundquist number $S_{\rm X}/a = 4\pi v_{\rm A} \Delta_0/\eta c^2 \sim 2.4$ and estimate for the reconnection rate $\sim S_{\rm X}^{-1/2} \sim (L_{\rm X} c / \eta)^{-1/2} \sim 0.32$.
The fact that our measured $\beta_{\rm rec} \sim 0.14$ is slightly smaller (relative to incompressible SP theory based on the local Lundquist number) may be due compressibility introducing a density difference between the charge-starved X-point and the (far) upstream $\sim {\rho_{\rm X}/\rho_0} \sim 0.2$ for which compressible SP would predict \cite{Parker1963,goodbred_22} $\beta_{\rm rec} \sim  (\rho_{\rm X} / \rho_{0})^{1/2}S_{\rm X}^{-1/2} \sim 0.14$.
Note that formally the incompressible limit cannot apply in the relativistic setting due to the sound speed becoming infinite, and indeed relativistic reconnection is strongly compressible  \cite{Grehan_2025}.
The density in the X-point for a given $\Delta_0$ is set by $a$, or equivalently by the slope of the magnetic field separatrix \cite{goodbred_22}.

\begin{figure}
    \centering
    \includegraphics[width=0.75\linewidth]{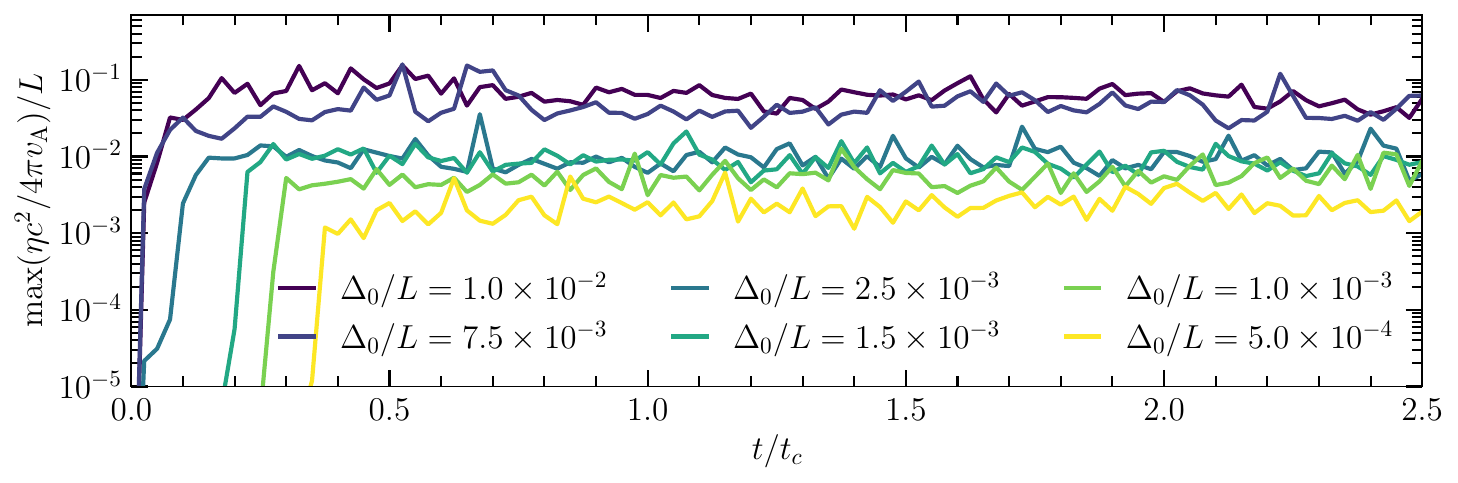}
    \caption{Time-evolution of the resistivity in an X-point, approximated as the maximum value in the domain. The resistivity depends on the skin depth $\sim \Delta_0/L$ (indicated by the colors of the lines) showing that $\eta |J_z|/B_0 \sim \beta_{\rm rec}$ remains constant.}
    \label{fig:eta_vs_delta}
\end{figure}

In Fig. \ref{fig:eta_vs_delta} we show the evolution of the maximum resistivity, ${\rm{max}}(\eta c^2/(4\pi v_{\rm A})) / L$, approximately corresponding to its value in X-points (see Fig. \ref{fig:1d_slice}), for a range of $\Delta_0/L$. 
As expected from its linear scaling with $\Delta_0$ (see Equation \ref{eq:1} in the main Letter), smaller $\Delta_0/L$ results in smaller resistivity (indicated by the colors of the lines). The reconnection rate and hence the non-ideal electric field $|\bm{E}^*|/B_0 \sim \eta |J_z|/B_0 \sim \beta_{\rm rec}$ (see Equation \ref{eq:1} in the main Letter) remain approximately constant for varying $\Delta_0$ (as long as $\Delta_0 \ll L)$. This means that the current $|J_z| \sim |\nabla \times \bm{B}| \sim B_0 / \Delta_0$ increases when the sheet thickness $\sim \Delta_0$ decreases and hence when $\eta$ decreases.

\begin{figure*}[htbp]
    \centering
    \begin{subfigure}{0.49\textwidth}
        \centering
        \includegraphics[width=\textwidth]{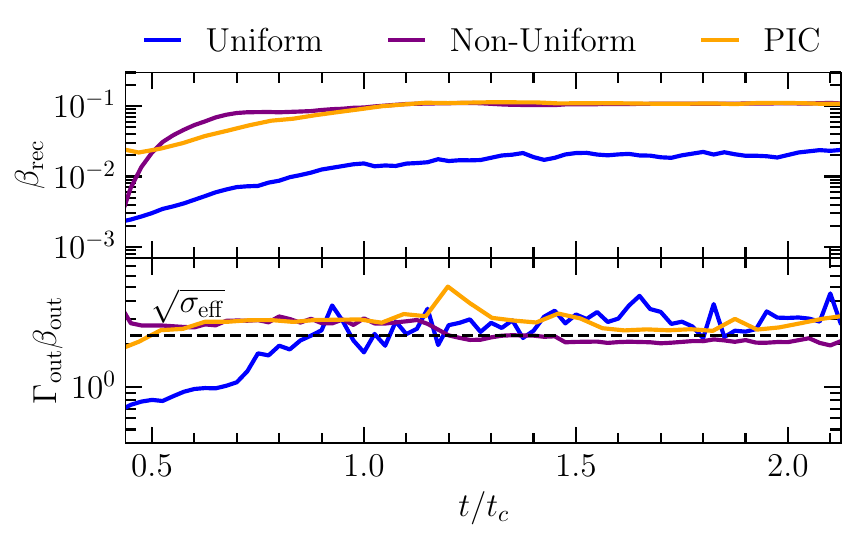}
        \caption{$B_{\rm g}/B_0 = 0.3$}
        \label{fig:bg1} 
    \end{subfigure}
    \hfill
    \begin{subfigure}{0.49\textwidth}
        \centering
        \includegraphics[width=\textwidth]{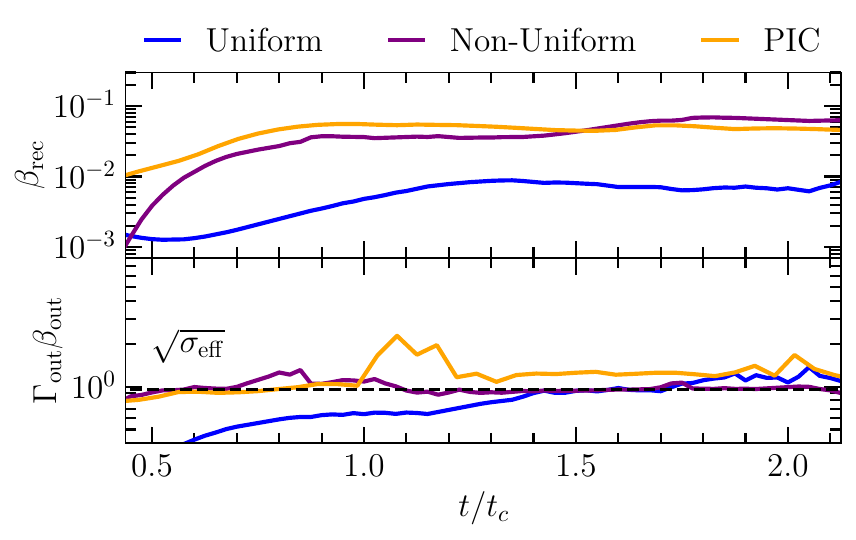}
        \caption{$B_{\rm g}/B_0 = 1.0$}
        \label{fig:bg3}
    \end{subfigure}
    \caption{Left panel: For guide field $B_{\rm g}/B_0=0.3$, we show the time evolution of the reconnection rate $\beta_{\rm{rec}}$ for RRMHD with uniform $\eta$ (blue), non-uniform $\eta$ (purple), and PIC (orange), showing that the non-uniform $\eta$ case matches the PIC reconnection rate $\sim 0.1$, while the uniform $\eta$ case shows the typical rate for RRMHD $\sim 0.02$. 
    Right panel: For guide field $B_{\rm g}/B_0=1$, we show the time evolution of the reconnection rate $\beta_{\rm{rec}}$ for RRMHD with uniform $\eta$ (blue), non-uniform $\eta$ (purple), and PIC (orange), showing that the non-uniform $\eta$ case matches the PIC reconnection rate $\gtrsim 0.05$, while the uniform $\eta$ case shows the typical rate for RRMHD $\sim 0.01$. Bottom panels: Time evolution of the four velocity of the outflow out of the X-point, which becomes Alfv\'{e}nic as an asymptotic steady state is reached for both guide field strengths. The value expected for an Alfv\'{e}nic outflow ($\Gamma_{\rm out}\beta_{\rm out} = \sqrt{\sigma_{\rm eff}}$) is indicated by the black dashed line}
    \label{fig:bg}
\end{figure*}

\section{Magnetic reconnection with a guide field}

In Fig. \ref{fig:bg} we show the time evolution of the reconnection rate $\beta_{\rm rec}$ (top panels) and outflow speed $\Gamma_{\rm out} \beta_{\rm out}$ (bottom panels), where we define the effective magnetization taking into account the inertia due to the guide field as \cite{werner_17}:
\begin{equation}
    \sigma_{\rm eff} = \frac{B_0^2}{4\pi \rho_0 (1 + \frac{\hat{\gamma}}{\hat{\gamma}-1} \frac{p_0}{\rho_0c^2})c^2 + B_{\rm g}^2},
\end{equation}
with adiabatic index $\hat{\gamma}=4/3$ and guide magnetic field with magnitude $B_{\rm g}$. The resulting in-plane (perpendicular to the guide field) Alfv\'{e}nic outflow velocity is
\begin{equation}
    v_{\rm A, \perp} = \sqrt{\frac{\sigma_{\rm eff}}{1+\sigma_{\rm eff}}} c.
\end{equation}
The outflow reaches the expected value of $\Gamma_{\rm out}\beta_{\rm out}= \sqrt{\sigma_{\rm eff}}$ \cite{werner_17, Grehan_2025,Sironi_2025}. 
For guide field $B_{\rm g}/B_0=0.3$ (left) the rate for non-uniform $\eta$ (for fiducial $\Delta_0/L = 5 \times 10^{-4}$) matches the rate in PIC, $\beta_{\rm rec} \sim 0.1$, whereas for uniform $\eta$, the rate is smaller $\beta_{\rm rec} \sim 0.02$. 
For guide field $B_{\rm g}/B_0=1$ (right) the rate for non-uniform $\eta$ (for fiducial $\Delta_0/L = 5 \times 10^{-4}$) matches the rate in PIC, $\beta_{\rm rec} \sim 0.05-0.06$, whereas for uniform $\eta$, the rate is smaller $\beta_{\rm rec} \sim 0.01$. 
We attribute the reduction of the reconnection rate with increasing guide field strength to the fact that the increase in guide field brings the system closer to the incompressible limit, which in the uniform $\eta$ case indeed results in the $\beta_{\rm rec} \sim 0.01$ expected from the incompressible MHD result \cite{Parker1963,lyutikov_uzdensky_03,lyubarsky_05}.
The functional form of the resistivity (which does not explicitly depend on $B_{\rm g}$ \cite{Moran_2025}) and the results matching the PIC reconnection rate for $B_{\rm g}/B_0=0.3$ and $B_{\rm g}/B_0=1$ indicate that our effective resistivity works for any guide field strength.

\bibliography{arxiv}


\end{document}